\documentclass[12pt]{iopart}

\usepackage{graphicx}
\usepackage{amssymb}

\font\tenbg=cmmib10 at 10pt

\def \rvecphi{{\hbox{\tenbg\char'036}}}

\def \rvecOmega {{\hbox {\tenbg\char'012}}}

\begin{document}

\title[]{Rossby Wave Instability in Astrophysical Discs}

\author{R.V.E. Lovelace{\footnote{lovelace@astro.cornell.edu}} and M.M. Romanova{\footnote{romanova@astro.cornell.edu}}}

\address{Department of Astronomy, Cornell University, Ithaca, N.Y. 14853, 
  USA}

\date{}

\begin{abstract}

      A brief review is given of the  Rossby wave instability (RWI) in
 astrophysical discs.
  In non-self-gravitating discs, around for example a newly
forming stars, the instability can be
triggered by an axisymmetric bump at some radius $r_0$ in the disc surface mass-density.
    It gives rise to exponentially growing non-axisymmetric perturbation
[$\propto \exp({\rm i}m\phi)$, $m=1,2..$]  in the vicinity
of $r_0$ consisting of {\it anticyclonic} vortices.  These vortices
are regions of high pressure and consequently act to trap  dust particles
which in turn can facilitate  planetesimal growth in proto-planetary discs.
   The Rossby vortices in the discs around stars and black holes may
cause the observed quasi-periodic modulations of the disc's thermal emission.

\end{abstract}
\vspace{2pc}

\maketitle

\section{Introduction}
   
\begin{figure*}[b]
\centering
\includegraphics[width=7cm]{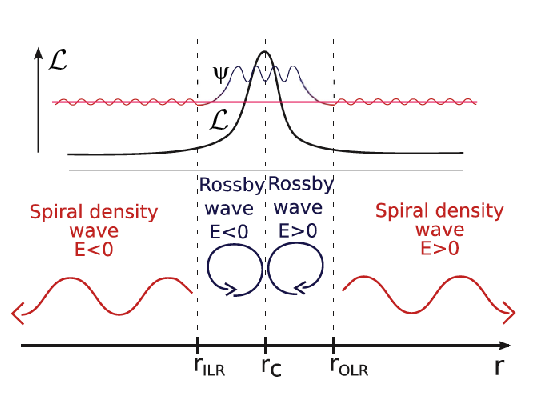}
\caption{Schematic view of the Rossby wave instability with
the two propagating regions for the Rossby waves, and in between
the evanescent regions close to the inner and outer Lindblad
resonant radii, ($r_{\rm ILR}$ and $r_{\rm OLR}$), respectively, adapted from Meheut \etal 2013).   The radii of these Lindblad
resonances are given by the equations $\omega = m \Omega(r_{\rm LR})\pm \kappa(r_{\rm LR})$, where $\kappa(r)$ is the radial epicyclic
frequency which is approximately equal to $\Omega(r)$ for a
Keplerian disc.  The corotation radius $r_{\rm C}$  is the
radius where $\omega = m \Omega(r_{\rm C})$.  }
\label{fig1}      
\end{figure*}

    The  theory of the  Rossby wave  instability (RWI) in accretion discs
was developed by Lovelace \etal (1999)
 and  Li \etal (2000) for thin Keplerian discs with negligible self-gravity  and earlier by  Lovelace and Hohlfeld (1978) for thin disc galaxies where the self-gravity may or may not be important and where the rotation is in general non-Keplerian.
      In the first case the instability can occur if there
is an axisymmetric bump (as a function of radius) in the inverse potential vorticity 
$$
{\cal L}(r) =
{\Sigma ~S^{2/\gamma} \over 2({\bf \nabla \times u})\cdot\hat{\bf z}}~,
\eqno(1)
$$
at some radius $r_0$, where $\Sigma$ is the surface mass density of
the disc, ${\bf u}\approx r\Omega(r)\hat{\rvecphi~}$ is the flow velocity of the disc,
$\Omega(r) \approx (GM_*/r^3)^{1/2} $ is the angular velocity 
of the flow (with $M_*$ the mass of the central star),    $S$ is the specific entropy
of the gas, and $\gamma$ is the specific heat ratio.   The approximations
involve the neglect of the relatively small radial pressure force (see Sec. 3).
Note that ${\cal L}$ is related to the inverse of the {\it vortensity}
which is defined as $({\bf \nabla \times u})_z/\Sigma$. A sketch of a bump in
${\cal L}(r)$ is shown in Figure \ref{fig1}.

   Rossby waves are important  in planetary atmospheres and oceans and
are also known as {\it planetary waves} (see  Rossby \etal 1939; 
Brekhovskikh and Goncharov 1993; Chelton and Schlax 1996;
Lindzen 2005).  These waves have a significant role in the transport of heat from  equatorial to polar regions of the Earth.   They may have a role in the formation
of the long-lived ($>300$ yr) Great Red Spot on Jupiter which is an anticyclonic vortex (e.g., Marcus 1993).
The Rossby waves have the notable property of having
the phase velocity opposite to the direction of motion of the atmosphere
or disc in the comoving frame of the fluid (Brekhovskikh
and Goncharov 1993; Lovelace \etal 1999).

   Section 2 summarizes the linear theory,  Sec. 3 describes dust
trapping in vortices,   Sec. 4 is on vortices formed at the radial margins
of the dead zone, Sec. 5 is on vortices due to planet induced
gap edges,  Sec. 6 is on the saturation and merging of vortices,
Sec. 7 is on the influence of a large-scale magnetic field, 
Sec. 8 is on vortices in self-gravitating discs, and Sec. 9 gives
a brief conclusion.

\section{Schr\"odinger-like equation for perturbation}

    Linearization of the Euler and continuity equations for a thin fluid
disc with perturbations proportional to $f(r)\exp({\rm i}m\phi
 -{\rm i }\omega t)$
(with azimuthal mode number $m=1, 2,..$ and angular frequency $\omega$) leads to a
 Schr\"odinger-like equation for the enthalpy perturbation $\psi=\delta p/\rho$,
$$
{d^2 \psi \over dr^2}= V_{\rm eff}(r)~\psi~.
\eqno(2)
$$
  The effective potential well $V_{\rm eff} (r)$ is closely related to ${\cal L}(r)$:  If the height of the  bump in ${\cal L}(r)$ is too small the potential well is shallow and there are no `bound Rossby wave states' in the well. On the other hand for a sufficiently large bump in ${\cal L}(r)$ the
potential $V_{\rm eff}$ is sufficiently deep to have a bound state.  The condition for there to be just one bound state allows one to solve for the 
imaginary part of the wave frequency, $\omega_i = \Im(\omega)$ which
is the growth rate of the instability (Lovelace et al. 1999).   
    For moderate strength bumps  (with fractional amplitudes 
$\Delta\Sigma/\Sigma \lesssim 0.2$),
the growth rates are of the order of $\omega_i = (0.1-0.2)\Omega(r_0)$.
The real part of the wave frequency $\omega_r =\Re(\omega)$ is
approxiamtely $m\Omega(r_0)$.   

\begin{figure*}
\centering
\includegraphics[width=10cm]{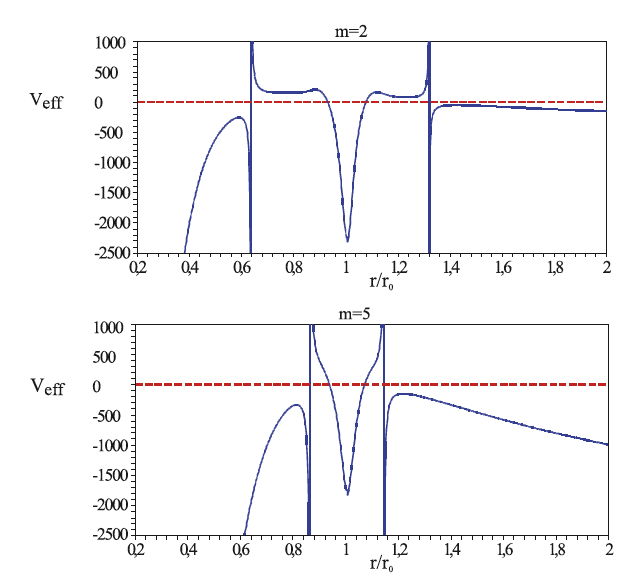}
\caption{Effective potential for a Gaussian surface density bump
of peak amplitude $\Delta \Sigma/\Sigma =0.2$ and width
$\Delta r/r = 0.05$ for 
$ m = 2$ (upper panel) and $m = 5$ (lower panel) adapted from
Meheut \etal (2013).  Waves can
propagate only in the regions where $V_{\rm eff}(r) < 0$.
  The large positive values of $V_{\rm eff}$ occur at the inner
and outer Lindblad resonance radii.}
\label{fig2}      
\end{figure*}

   A more complete analysis (Tagger 2001; Tsang and Lai 2008; Lai and
Tsang 2009) reveals that the Rossby wave is not completely trapped in the potential well $V_{\rm eff}$, but leaks outward across a forbidden region at
an outer Lindblad resonance (at $r_{\rm OLR}$ indicated in
Figure \ref{fig1}) and inward across another forbidden
region at an inner Lindblad resonance (at $r_{\rm ILR}$). 
 Once the waves cross the forbidden regions they propagate as spiral density wave.
 The  full expression for the
effective potential for a thin {\it homentropic} ($S=$ const)  disc is
$$
V_{\rm eff} = {2m \Omega \over r (\Delta \omega)}
{d\over dr}\left[\ln\left({\Omega \Sigma 
\over \kappa^2-(\Delta \omega)^2}\right)\right]+{m^2 \over r^2}
+{\kappa^2-(\Delta \omega)^2 \over c_s^2}~,
\eqno(3)
$$
where $\Delta \omega \equiv \omega-m\Omega$ is the Doppler
shifted wave frequency in the reference frame moving with the disc
matter,  $c_s$ is the sound speed in the disc, and
 $\kappa$  is the radial epicyclic angular frequency,
with $\kappa^2=r^{-3}d\ell^2/dr$ and $\ell=ru_\phi$ the
specific angular momentum (Meheut \etal 2013).
      Figure \ref{fig2} shows the effective potential for sample
cases.   
    Note  that the inward
propagating waves with $\omega_r < m \Omega(r)$
 have negative energy ($E<0$) whereas the
outward propagating waves with $\omega_r > m \Omega(r)$ 
have positive energy ($E>0$) (e.g., Meheut \etal 2013).

    The Rossby wave instability is analogous to the  {\it plane parallel
shear instability} which can be interpreted in terms of counter-propagating Rossby waves on either side of the resonance layer ($r_0$) and  the over-reflection of waves from this resonance layer (Lindzen 1988;  Harnik and Heifetz 2007).  The analogy remains to 
be fully investigated.
An interpretation of the Rossby instability
as arising from the interaction of two  ``edge waves'' on either side
of the bump in ${\cal L}(r)$ is given by Umurhan (2010) who points
out the analogy of the RWI with the {\it diochotron} instability 
(Buneman 1957; Knauer 1966) of
rotating, non-neutral plasmas in a uniform magnetic field for cases where the plasma is annular with
both inner and outer radii so that there  are two edge waves.

    The Rossby wave instability occurs because of the local wave trapping
in a disc.   It is related to the Papaloizou and Pringle (1984, 1985)
instability where the wave is trapped between the inner and outer radii of
a disc or torus.

\begin{figure*}
\centering
\includegraphics[width=12cm]{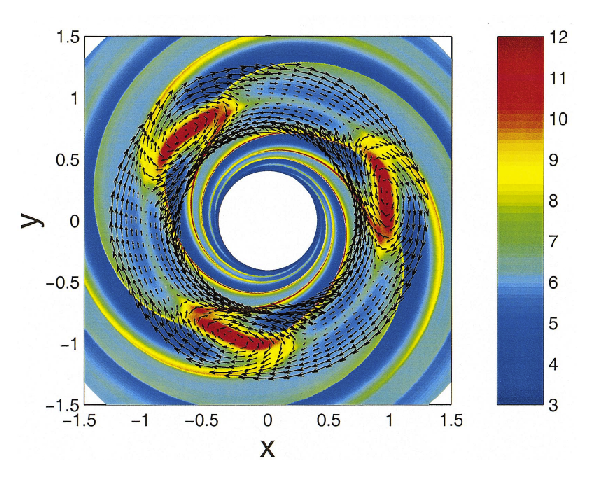}
\caption{2D hydrodynamic simulations of Rossby vortices in a disk
adapted from Li \etal (2001)
 for $m=3$.
Pressure is color-coded (in units of $10^{-3}p_0$). 
Arrows indicate the flow pattern near $r_0$ in a comoving frame moving with
velocity $u_\phi(r_0)$.  The vortices are 
{\it anticyclonic}, enclosing high-pressure regions. 
Large-scale spirals are produced as well, in connection with the vortices. }
\label{fig3}      
\end{figure*}

   Two-dimensional hydrodynamic simulations of the RWI
instability in discs have been done by many groups beginning
with Li \etal (2001).   Figure \ref{fig3} from Li \etal (2001)   shows
an example of the $m=3$ vortices formed by the RWI.   
    Three-dimensional simulations of
the instability have been done  by Meheut \etal (2010, 2012a).
     The behavior of the Rossby wave instability is not
dramatically altered in going from the 2D to 3D hydrodynamics.  In
3D there are vertical ($z$) motions associated with the vorticies, but
the magnitudes of the vertical velocities are small compared with
the horizontal perturbation velocities.    Earlier 3D simulations
of vortices in stratified proto-planetary discs by
Barranco and Marcus (2005) gave a more complicated picture
of the vortex dynamics.

The theory of the instability in three dimensions 
has been developed by Meheut \etal (2012b) and Lin (2012).    
A useful set of references can
be found in the proceedings  of a recent meeting  on vorticies and
dust dynamics in proto-planetary discs edited by
Barge and Jorda (2013).

\section{Trapping of dust accretion disc vortices}

    Barge and Sommeria (1995) first pointed out that anticyclonic vortices in an accretion disc can act to concentrate dust particles in the flow and that this would
facilitate the formation of planetesimals.  This subject has been 
further developed in many subsequent papers (e.g.,  Tanga \etal 1996;
Bracco \etal 1999; Godon and Livio 2000; Johanssen \etal 2004;
Heng and Kenyon 2010). 
A qualitative explanation
is simple.  
     First note that in an axisymmetric proto-stellar disc the  radial force balance
for the gas component  is $-u_{\phi \rm gas}^2 /r = -(dp/dr)/\rho - GM_*/r^2$, where $u_{\phi\rm gas}$ is gas velocity, $p$ is its pressure,  $\rho$ is its density, and $M_*$ is the mass of the central star.
The pressure gradient term can be estimated as $(dp/dr)/\rho \approx - c_s^2/r$,
with $c_s = (p/\rho)^{1/2}$ the sound speed of the gas.  
     Thus the gas velocity 
is slightly sub-Keplerian, $u_{\phi\rm gas} \approx v_{\rm K} [1-(c_s/v_{\rm K})^2/2]$, where
$v_{\rm K} \equiv (GM_*/r)^{1/2}$ is the Keplerian velocity and $c_s/v_{\rm K} \ll 1 $ for
proto-stellar discs (where typical value is $c_s/v_{\rm K} =0.05 $).
    On the other hand the dust component of the proto-stellar disc is not affected by
a pressure gradient and consequently the dust velocity is
Keplerian $u_{\phi\rm dust} = v_{\rm K}$.   
       The dust particles move slightly faster than the gas and consequently  
lose angular momentum by the drag from their motion through the gas.
     Dust particles of sizes $0.01 - 1$ cm 
inspiral to the proto-star faster than the gas for commonly assumed
disc viscosities.   A rigorous 3D analysis of the dust motion in axisymmetric discs is given by Takeuchi and Lin (2002).
  
\begin{figure*}
\centering
\includegraphics[width=8cm]{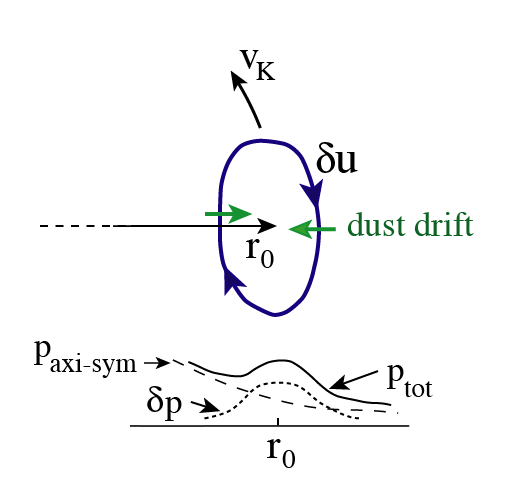}
\caption{Sketch of an anticyclonic vortex centered at radius 
$r_0$ in a Keplerian disc at some instant of time.  The velocity
perturbation of the vortex is $\delta {\bf u}$.  The lower part
of the figure sketches the radial pressure variation through the
vortex center.    }
\label{fig4}      
\end{figure*}
  
       Consider a vortex of small radial and
  azimuthal  extent formed by some process in a disc as sketched in Figure \ref{fig4}.
 It has  velocity perturbation $\delta {\bf u}$ and pressure perturbation $\delta p$.
   In a reference frame rotating with angular rate $\Omega(r_0)~(\approx
   v_{\rm K}(r_0)/r_0)$, the flow
 approximately satisfies the geostropic equation  $(\nabla \delta p)/\rho
 = 2\delta {\bf u} \times \rvecOmega$, which is a balance of the pressure gradient
 and Coriolis forces.   An {\it anticyclonic vortex}, with vorticity ${\bf \nabla \times \delta u}$
 oppositely directed to $\rvecOmega$,  has a maximum of $\delta p$ at the vortex center.
 (A  cyclonic vortex has a pressure minimum at its center.)   A dust particle in
 the outer part of the vortex   moves faster than the gas and drifts radially inward.   
   On the other hand a dust particle in the inner part may move slower than the
 gas with the result that it drifts radially outward.  
   For dust trapping the vortex must be strong enough for the radial
pressure gradient to be positive for an interval inside $r_0$.  This gives
the condition $|\delta {\bf u}| \gtrsim c_s^2/2v_{\rm K}$, which is readily
satisfied by the vortices arising from the RWI.

      Dust trapping in vorticies in 3D
disc simulations has been studied by modeling the dust component as a
separate pressureless fluid coupled to the gas by collisions 
(Meheut \etal 2012c)
for dust particle sizes of $0.1-5$ cm.
   The dust to gas mass fraction, initially $10^{-2}$, was  found to increase by a factor of order $10^2$ in the centers of the  vortices (Meheut \etal 2012c).
    At this mass fraction the dust has a significant back reaction 
on the gas dynamics.

\section{Vortices at the radial margins of the dead zone in disc}

    A   bump in ${\cal L}(r)$ or $\Sigma(r)$  may  arise 
at the radial boundaries of the ``dead zone''
(Varni\`ere and Tagger 2006; Lyra \etal 2009; Crespe \etal 2011).  
   This zone can arise from the suppression of 
the  magneto-rotational instability (MRI, Balbus and Hawley 1991)  by low ionization  inside the disc (Gammie 1996).  The MRI is a small length-scale  instability of conducting Keplerian discs in the presence of
 a  weak magnetic field where the magnetic pressure is less
 than the thermal pressure in the disc.   The instability is found
 to give rise to  magnetic turbulence with length-scales of the order of the disc thickness.  
     In turn this turbulence provides an effective 
 viscosity which is commonly thought to cause the accretion of matter and the outward  transport of angular momentum in
 astrophysical discs (Balbus and Hawley 1991).
    Note however the possibility  that the turbulent viscosity  arises in  hydrodynamic (non-magnetized) discs owing to the nonlinear instability in the very large Reynolds number,
Rayleigh-stable Keplerian discs (specific angular momentum 
increasing outward) ( Paoletti \etal 2012;  Bisnovatyi-Kogan and Lovelace 2001;  Zeldovitch 1981).
    The bump in ${\cal L}(r)$  can give rise to 
exponential growth of non-axisymmetric perturbations in the vicinity
of $r_0$ consisting of {\it anticyclonic} vortices.
    In most studies the dead zone has been modeled as radial interval of the disc
with a significantly reduced turbulent viscosity compared with
the outer and inner parts of the disc.   This is due to the reduced action of
the MRI.
However,  recently, Lyra and Mac Low (2012) carried out global 3D 
magneto-hydrodynamic (MHD) simulations of
the full problem by modeling the dead zone as a region of the disc with
significantly lower resistivity.  They find that the RWI develops on the inner edge of the dead zone.

\section{Vortices due to planet induced gap edges}

   The interactions of a planet with 
a proto-stellar disc has been studied theoretically and with
2D and 3D hydrodynamic simulations by many authors primarily because
of its importance to planet migration, that is, the slow inward or outward
radial motion of the planet due to transfer of angular momentum between
the planet and the disc.  The dependence of the migration rate and the
influence of a planet on the disc (e.g., the formation of a gap in the disc) 
depends of course on the planet's mass,  and it was  thought 
that the formation of a gap in the disc occurs only 
 for planets more massive than about one-tenth of Jupiter's mass (Lin and
 Papaloizou 1993).
   But the gap formation is sensitive also to
 the nonlinear steepening and shock formation of the spiral
density waves generated by the planet (Rafikov 2002)  as
well as the  turbulent viscosity of the
disc which damps the density waves (Li \etal 2009). 
    For sufficiently low viscosities, the shocks formed by
the spiral density waves  at the gap edges give rise to
maxima  in the inverse vortensity ${\cal L}(r)$  (Koller \etal 2003;
Li \etal 2005; de Val-Borro \etal 2007;  Lin and Papaloizzou 2010).  This triggers the Rossby wave instability.    
    Starting from axisymmetric conditions the simulations show the growth
and saturation of several vortices ($m=3-5$) which later merge to form
a single vortex.

   Recent high resolution 3D hydrodynamic simulations by Zhu \etal (2014)
indicate that a gap may be formed over a sufficiently 
long period for $M_p $ larger than ten times the earth mass.
       In this work the gas component of the disc was modeled with the grid
based code {\it Athena}, and the dust component with  Lagrangian  particles
for a wide range of dust particle sizes ($0.01~-~10^2$ cm).  
  The dust surface mass density in the Rossby vorticies 
is found to increase by a factor of $10^2$ for dust particles with  
$\Omega_{\rm K} t_s \sim 1$, where $\Omega_{\rm K}$
is the Keplerian rotation rate of the disc at $r_0$,  and $t_s$ is the stopping time of the dust particle motion through the gas 
(proportional to the particle's radius)  (Zhu \etal 2014). 
Note that in the work of Meheut  \etal (2012c), where the dust component was treated
as a separate fluid,  a comparable increase in the dust density was
found.   Analytic solutions for the dust distribution
in disc vortices have been developed by  Birnstiel \etal (2013)
and Lyra and Lin (2013).

    Recent observations by van der Marel \etal (2013) of the disc around the star  Ophiuchus IRS 48
with the Atacama Large Millimeter Array (ALMA) have revealed
a high-contrast crescent-shaped emission region in the $0.44$
mm wavelength from one side of the star indicating
a concentration in this region of millimeter size dust grains.
    The contrast ratio between the peak emission and the opposite
side of the disc is $\sim 10^2$.
At the same time observations at different wavelengths indicate
that both the gas and the micrometer-sized dust is distributed uniformly
in the disc around the star (van der Merel \etal 2013).
     The asymmetric dust emission of Oph IRS 48 is modeled as
a single anticyclonic  Rossby vortex ($m=1$) excited at the outer gap edge of a $\sim 10$ Jupiter mass planet orbiting the star at $20$ AU
(AU is  the Sun-Earth distance, $1.5\times 10^{13}$ cm) (van der Marel
\etal 2013).    The center of the vortex or dust trap is estimated as
$63$ AU and it is on the line through the planet and star.

\section{Saturation and merging of vorticies}

   Starting from axisymmetric conditions, a bump in ${\cal L}(r)$
of  amplitude larger than a critical value will lead to the exponential growth of a Rossby wave  for a particular azimuthal mode number (typically, $m= 3-5$) (Lovelace \etal 1999).
   The exponential growth is predicted to cease at the time when
the circulation or trapping time of a fluid particle in one of the vortices,
$\tau_T$, is of the order of the  growth-time of the instability,
$\omega_i^{-1}$ (Lovelace \etal 2009).
  This is because the axisymmetric fluid motion assumed in the instability calculation (Lovelace \etal 1999) is inapplicable for times $\gtrsim \tau_T$.
     An analogous saturation conditions is well-known in plasma 
physics where it was first used to explain the saturation of the
two-stream instability (O'Neil 1965; Krall and Trivelpiece 1973).
      A systematic  study of the Rossby wave saturation with 2D and a sample 3D hydrodynamic simulations finds that the exponential growth saturates when $\tau_T \lesssim \omega_i^{-1}$ with
 $\tau_T \approx 2/|\omega_v|$ where $\omega_v =|{\bf \nabla \times
 \delta u}|$ is the vorticity of the Rossby vortex (Meheut \etal 2013).  
     The saturation is found to occur in about $5 - 10$ orbital periods  at
 the radius of the bump (Meheut \etal 2013).

     An axisymmetric bump in the inverse vortensity ${\cal L}(r)$ may be maintained at the edge of a disc's dead zone  or at the edges of a gap formed in the disc by a planet.       The maintenance of the bump can allow the continuous driving of the Rossby vortices for the lifetime
 of the disc.
     In the absence of a mechanism
 for maintaining the bump, long-time 3D  hydrodynamic simulations show  that that the Rossby vortices persist for times of the order of $10^2$  orbital
 periods at the radius of the bump (Meheut \etal 2012a).
 For longer times the bump in ${\cal L}$ is spread
 out in radius and reduced in amplitude below the instability threshold.
   The vortices do not show measurable radial migration.
 
     The long-time simulations show that an initial  multiple vortex
 pattern ($m=3-5$) evolves by vortex merging to give a 
 single vortex ($m=1$) (Meheut \etal 2012a).
    A detailed theoretical analysis of the stability of a single elliptical Kida  vortex model (1981)  has been carried out by Leseur and Papaloizou (2009) with the result such vortices are unstable except for azimuthally elongated vortices
which have azimuthal half-widths ($r_0\Delta \phi$) longer than $4$ times the radial half-width ($\Delta r$) but with this ratio less than $6$.

\section{Influence of a large-scale magnetic field on vortices}
    
            The influence of a large-scale  magnetic
field on the Rossby wave instability has been investigated by
 Tagger and collaborators (Tagger and Pellat 1999;
 Tagger and Varni\`ere 2006; Tagger and Melia 2006)  and
 others (Yu and Li 2009; Yu and Lai 2013). 
   Strong, kilo-Gauss,   dipole and higher multipole moment
magnetic fields are observed in proto-stars (e.g., Yang and Johns-Krull 2011).
  The stellar magnetic field penetrates   the proto-stellar
disc to different radial extents depending mainly on the stellar field 
strength and the disc accretion rate (Lii \etal 2012).   This large-scale
field in the inner disc can drive the high velocity jets observed owing
to the gradient of the wound-up toroidal magnetic fields (Lii \etal 2012).
  At the same time the toroidal magnetic
field collimates the jet.  Large-scale magnetic fields are also thought
to be responsible for the relativistic jets observed to
come from accretion discs around
black holes ranging from stellar mass to supermassive ($\sim 10^{10}$
solar mass).  In the presence of a large-scale, vertical magnetic field
$B_z(r)$,  Tagger and collaborators find that the inverse vortensity of equation (1) may be
superceded by the quantity $(B_z/\Sigma)^2 {\cal L}$ depending
on the value of $|B_z|/\Sigma$.

      A distinguishing aspect of the 
magnetic RWI is the prediction of the outflow of energy and
angular momentum via Alfv\'en waves from the vortices into the disc's corona 
(Tagger and Pellat 1999;  Tagger and Varni\`ere 2006).
  The outflows from the rotating vortices could explain the
quasi-periodic oscillations (QPOs) observed in the X-ray emissions of
low-mass X-ray binaries (van der Klis 2006).      
          A further promising application of the magnetic RWI is to
modeling the quasi-periodic ($\sim 20$ min.)  near infra-red
and X-ray emission from material orbiting the super massive black 
hole Sagittarius A$^*$ ($M\sim 3\times 10^6$ solar masses) in the center of our Galaxy (Tagger and Melia 2006).
   The authors propose that  low angular momentum clouds episodically
fall into the vicinity of the black hole and the gas becomes
trapped and circularized by viscosity at $\sim 30$ Schwartzchild
radii ($r_S=2GM/c^2 \sim 10^{12}$ cm) from the black hole.   This event would naturally produce a pronounced bump in the disc surface density and ${\cal L}(r)$,
strong Rossby wave growth at the trapping radius, and 
magnetically driven outflows with a quasi-period similar to that
observed.      Note that Vincent \etal (2013) find quasi-periodic oscillations
in the thermal emission from the inner regions of {\it non-magnetized} discs around black holes.  They find the RWI in  from 2D and 3D simulations using
the {\it AMRVAC} code.

            A different type of Rossby wave instability was 
found by Lovelace \etal (2009) in the strongly non-Keplerian magnetized
region of a disk where $d\Omega/dr >0$.  For a Keplerian disc,
$d\Omega/dr = -(3/2)\Omega/r$.
      This region occurs when an accretion disc encounters the magnetosphere of a  slowly rotating star (Romanova \etal 2008).   This Rossby mode was proposed as an explanation (Lovelace \etal 2009) of the observed twin kilo-Hertz QPOs (van der Klis 2006).   Evidence for the occurrence of
 this Rossby mode with $m=2$ and $d\Omega/dr>0$ has been found in the power spectral density plots of the global 3D MHD simulations of waves in discs around rotating
 magnetized stars with misaligned dipole magnetic fields using the
 {\it Cubed Sphere} code (Romanova \etal 2013).

\section{Vortices in self-graviating discs}

    The self-gravity of disc can have an important influence on
the Rossby wave instability in proto-planetary discs, black hole
discs, as well as disc galaxies.   The  RWI, earlier
termed the  ``negative
mass instability'' (Lovelace and Hohlfeld 1978), was in fact developed for spiral or disc
galaxies which are self-gravitating systems of stars and gas.
      The rotation curves the disc galaxies are {\it not} Keplerian, but
 ``flat''  because they have $u_\phi \approx $ const for a large range
 of radial distances outside their central nuclear regions.
     For disc galaxies ${\cal L} \rightarrow \sigma \Omega/\kappa^2$
 (Lovelace and Hohlfeld 1978), 
where $\kappa$  is the radial epicyclic angular frequency given
below equation 3.
   The RWI in disc galaxies consisting predominanly of
stars has been observed and studied in many different
cases using $N-$body simulations (Sellwood and Kahn 1991;
Sellwood 2012; Sellwood 2013).  
       Only a small bump or depression in ${\cal L}(r)$ leads
 to the growth of a global disc mode (Sellwood 2013).
   When the  self-gravity of the disc is strong, the RWI
is observed to occur at radii where ${\cal L}$ is a {\it minimum}
rather than a maximum  (Sellwood and Kahn 1991;
Sellwood 2012, 2013)
as predicted by Lovelace and Hohlfeld (1978).

       In proto-stellar discs, the influence of the self-gravity of the gas and dust on vortices generated by giant Jupiter-mass scale planets  have been studied in several recent simulation studies (Lyra \etal 2009;  Lin and Papaloizou 2011).
   Lyra \etal (2009) use a thin disc, 2D model for the gas
and dust with the gas described by hydrodynamic equations and
the dust described by a large number of Lagrangian particles
(typically $10^5$) using the {\it Pencil} code.   The dust and gas are coupled by an analytic drag force model which allows for a wide range of particle sizes and gas conditions.   The self-gravity is calculated using a Fourier transform method.
Significant dust accumulation is found to occur at the Lagrange 
points of the planet as well as in the vortices generated at the 
edges of the planet induced gap in the disc.
    The analysis of  Lin and Papaloizou (2011) is also for  thin discs using analytic methods and 2D hydrodynamic simulations with the {\it Fargo} code.  
The vortices studied were
 those arising from the planet induced bumps in ${\cal L}(r)$ at the gap
edges.  The analytic theory and simulations show that the Rossby instability is stabilized for low mode numbers ($m$) as the disc mass  is increased.

      A simpler problem is treated by Lovelace and Hohlfeld (2013) where an
analytic calculation is made of the stability of a bump or depression 
in ${\cal L}(r)$ in a thin Keplerian disc in the absence of planets.  The results
are in agreement with Lovelace and Hohlfeld (1978).
 The key quantities
determining the stability/instability are found to be: (1) The parameters of the bump
(or depression) in the disc surface density. (2) The Toomre
$Q=\kappa c_s(\pi G \Sigma)^{-1}$ parameter of the disc with $c_s$
is the sound speed in the disc (Toomre 1964; Safronov 1960) where
a non-self-gravitating disc
has $Q\gg1$.
And, (3) the dimensionless azimuthal wavenumber of the 
perturbation $\overline{k}_\phi =mQh/r_0$, where $h$ is the half-thickness of the disc.   
      For discs stable to axisymmetric perturbations ($Q>1$),  the self-gravity 
has a significant role   for $\overline{k}_\phi  <  \pi/2$ or $m<(\pi/2) (r_0/h)Q^{-1}$;   instability may occur  for a depression or groove in the surface density if $Q\lesssim 2$.
  For  $\overline{k}_\phi  >  \pi/2$ the self-gravity is not important, and
instability may occur at a bump in the surface density.
Thus, for all mode numbers $m \ge 1$, the self-gravity is unimportant
for $Q > (\pi/2)( r_0/h)$.   
    This review does not consider the instability of self-gravitating gas discs for conditions where the gas cooling is important (Gammie 2001).
     
 \section{Conclusion}
 
    Recent research on vortices in astrophysical discs is evolving  particularly rapidly
owing  to the completion of the ALMA infra-red telescope array which 
allows for the first time the mapping of the distribution of gas and dust  grains
in the discs of nearby proto-planetary systems.   
  Observations by van der Marel \etal (2013) with ALMA of a crescent shaped dust 
concentration in the disc of a proto-star is most simply interpreted as
a large mass of dust trapped in the high pressure region of an anticyclonic vortex.
  At the same time increasingly sophisticated codes have been
developed  for simulating  the 3D flow
of gas and dust in discs (e.g., Zhu \etal 2014).

\ack

   We thank Professor Michael Mond of Ben-Gurion University
and Professor Alexander Oron of the Technion   for the
invitation to the meeting {\it Bifurcations and Instabilities in 
Fluid Mechanics} held at the Technion in Haifa Israel in July 2013.
Further, we thank Dr. Orkan  Umurhan for valuable discussions on
Rossby vortices.

\section*{References}
\begin{harvard}
\item[] Balbus, S.A., and Hawley, J.F. 1991, ApJ, 376, 214

\item[]  Barge, P.,  and Jorda, L. (eds.) 2013,  {\it Instabilities and Structures in Proto-Planetary Disks}, EPJ Web of Conferences 46, 2013
 
\item[]  Barge, P.,  and Sommeria, J. 1995, A\&A, 295, L1

\item[] Barranco, J.A., and Marcus, P.S. 2005, ApJ, 623, 1157
 
\item[]  Birnstiel, T.,  Dullemond, C.P.,  and Pinilla, P. 2013, A\&A, 550, L8

\item[] Bisnovatyi-Kogan, G.S., and Lovelace, R.V.E. 2001, New
Astronomy Reviews, 45, 663
 
\item[]  Bracco, A., Chavanis, P.H., Provenzale, A.,  and Spiegel, E.A. 1999, Phys. Fluids, 11, 2280
 
\item[] Brekhovskikh, L. M.,  and Goncharov, V. 1993, {\it Mechanics of Continua and Wave Dynamics} (Berlin: Springer), pp 246-252

\item[] Buneman, O. 1957, J. Electronics and Control, 3, 507

\item[] Chelton, D.B., and Schlax, M.G. 1996, Science, 272, 234

\item[]  Crespe E., Gonzalez J.-F.,  and  Arena S. E., 2011, in SF2A-2011: Proceedings of the Annual meeting of the French Society of Astronomy and Astrophysics, G. Alecian, K. Belkacem, R. Samadi,  and D. Valls-Gabaud, eds., (Societe Francaise d'Astronomie et d'Astrophysique (SF2A)),  pp. 469-473
 
\item[]  de Val-Borro, M.,  Artymowicz, P., D'Angelo, G.,  and Peplinski, A. 2007, A\&A, 471, 1043
 
 \item[] Gammie, C.F. 1996, ApJ, 457, 355
 
\item[]  Gammie, C.F. 2001, ApJ, 553, 174
 
\item[] Godon, P.,  and Livio, M. 2000, ApJ, 537, 396

\item[] Harnik, N., and Heifetz, E. 2007, J. Atmospheric Sciences,
64, 2238

\item[] Heng, K.,  and Kenyon, S.J. 2010, MNRAS, 408, 1476

\item[] Johanssen, A., Andersen, A.C.,  and Brandenburg, A. 2004, A\&A, 417, 361

\item[]  Kida, S. 1981, Phys. Soc. Japan J., 50, 3517

\item[] Knauer, W. 1966, J. Appl. Phys., 37, 602
 
\item[]  Koller, J., Li H.,  and Lin D. N. C.  2003, ApJ, 596, L91

\item[] Krall, N., and Trivelpiece A. 1973, {\it Principles of Plasma Physics}, McGraw-Hill,
New York, p. 536
 
 \item[] Lai, D.,  and Tsang ,D., 2009, MNRAS, 393, 979
 
\item[]  Leseur, G.,  and Papaloizou, J.C.B. 2009, A\&A, 498, 1
 
 \item[] Li, H., Colgate, S. A.,  Wendroff, B.,  and  Liska, R. 2001, ApJ, 551, 874
 
\item[]  Li, H., Finn, J.M., Lovelace, R.V.E.,  and  Colgate, S.A. 2000, ApJ, 533, 1023
 
\item[]  Li, H., Li, S., Koller, J. Wendroff, B.B.,  Liska, R., Orban, C.M., Liang, E.P.T.,  and Lin, D. N. C. 2005, ApJ, 624, 2003
 
 \item[] Li, H., Lubow, S.H., Li, S.,  and Lin, D.N.C. 2009, ApJ, 690, L52
 
\item[]  Lii, P., Romanova, M.M.,  and Lovelace, R.V.E. 2012, MNRAS, 420, 2020
 
\item[]  Lin, D.N.C.,  and Papaloizou, J.C.B. 1993, in {\it Protostars and Planets III}, eds.
E.H. Levy and J.I. Lunine, (Univ. of Arizona Press: Tuscon), 749
 
\item[]  Lin, M.-K., 2012, ApJ, 754, 21
 
\item[]  Lin, M.-K.,  and Papaloizou, J.C.B. 2010, MNRAS, 405, 1473
 
\item[]  Lin, M.-K.,  and Papaloizou, J. C. B. 2011, MNRAS, 415, 1426

\item[] Lindzen, R.S. 1988, Pure Appl. Geophys., 126, 103

\item[] Lindzen, R.S. 2005, {\it Dynamics in Atmospheric Physics}
(Cambridge: Cambridge Univ. Press), pp 222-233
 
\item[]  Lovelace, R.V.E., Li, H., Colgate, S.A.,  and Nelson, A.F. 1999, ApJ, 513, 805
 
 \item[] Lovelace, R.V.E.,  and Hohlfeld, R.G. 1978, ApJ, 221, 51
 
\item[]  Lovelace, R.V.E.,  and Hohlfeld, R.G. 2013, MNRAS, 429,  529
 
\item[]  Lovelace R. V. E., Turner L.,  and Romanova M. M., 2009, ApJ, 701, 225
 
\item[] Lyra, W., Johansen, A., Zsom, A., Klahr, H.,   and Piskunov, N. 2009, A\&A, 497, 869

\item[]  Lyra, W.,  and Lin, M.-K. 2013, ApJ, 775, article id. 17, pp. 1-10 (arXiv:1307.377)
 
\item[]  Lyra, W.,  and Mac Low, M.M. 2012, ApJ, 756, 62

\item[] Marcus, P.S. 1993, Ann. Rev. Astron. and Astrophys., 31, 523
 
\item[]  Meheut, H., Casse, F., Varni\`ere, P.,  and Tagger, M. 2010, A\&A, 516,  A31
 
\item[] Meheut, H., Keppens, R., Cassee, F.,  and Benz, W. 2012a, A\&A, 542,  A9

\item[] Meheut, H., Yu, C.,  and Lai, D. 2012b,  MNRAS, 422, 2399

\item[]  Meheut, H., Meliani, Z., Varniere, P.,  and Benz, W. 2012c,  A\&A, 545. A134

\item[]  Meheut, H., Lovelace, R.V.E.,  and Lai, D. 2013, MNRAS, 430, 1988

\item[]  O'Neil, T. 1965, Phys. Fluids, 8, 2255

\item[] Paolettti, M.S., van Gils, D.P.M., Dubrulle, B., Sun, C., Lohse, D.,
and Lathrop, D.P. 2012, A\&A, 547, A64
 
\item[]  Papaloizou, J. C. B.,  and Pringle, J. E. 1984, MNRAS, 208, 721; ------- 1985, MNRAS, 213, 799
 
\item[]  Rafikov, R.R. 2002, ApJ, 572, 566
 
\item[]  Romanova, M. M., Kulkarni, A. K.,  and Lovelace, R. V. E. 2008, ApJ, 673, L171
 
\item[] Romanova, M. M.,  Ustyugova, G. V., Koldoba, A. V.,  and Lovelace, R. V. E. 2013, MNRAS, 430, 699

\item[] Rossby C.-G. and Collaborators 1939, Journal of Marine
Research, 2, 38

\item[]  Safronov, V.S. 1960, Ann. Astrophys., 23, 982
 
\item[] Sellwood, J.A.,  and Kahn, F.D. 1991, MNRAS, 250, 278

\item[]  Sellwood, J.A. 2012, ApJ, 751, 44
 
\item[]  Sellwood, J.A. 2013, Chap. 18, ``Dynamics of Disks and Warps,'' in {\it Planets, Stars and Stellar Systems Vol. 5}, eds: Oswalt, T.  D.  and Gilmore, G., (ISBN 978-94-007-5611-3. Springer Science+Business Media Dordrecht), p. 923 (arXiv: 1006.4855)
 
\item[]  Tagger, M., 2001, A\&A, 380, 750
 
\item[]  Tagger, M.,  and Melia, F. 2006, ApJL, 636, L33
 
\item[] Tagger, M.,  and Pellat, R. 1999,  A\&A, 349, 1003

\item[]  Tagger, M.,  and Varni\`ere, P. 2006,  ApJ, 652, 1457
 
\item[]  Takeuchi, T.,  and Lin, D.N.C. 2002, ApJ, 581, 1344
 
\item[]  Tanga, P., Babiano, A., Dubrulle, B., \& Provenzale, A. 1996, Icarus, 121, 158
 
\item[]  Toomre, A. 1964, ApJ, 139, 1217
 
\item[]  Tsang, D.,  and Lai, D., 2008, MNRAS, 387, 446
 
\item[]  Umurhan, O.M. 2010, A\&A, 521, A25
 
\item[]  van der Klis, M. 2006, in Compact Stellar X-Ray Sources, ed. W. H. G. Lewin  and M. van der Klis (Cambridge: Cambridge Univ. Press), 39
 
\item[]  van der Marel, N.,  van Dishoeck, E.F.,  Bruderer, S., Birnstiel, T.,  Pinilla, P.,  Dullemond, C.P., van Kempen, T.A.,  Schmalzl, M., Brown, J.M., Herczeg, G.J.,  Mathews, G.S.,  and Geers, V. 2013, Science, 340, 1199, (arXiv:1306.176)

\item[]  Varni\`ere, P.,  and Tagger, M. 2006, A\&A, 446, L13

\item[] Vincent, F.H., Meheut, H., Varni\`ere, P., and Paumard, T. 2013,
A\&A, 551, A54
 
\item[]  Yang H.,  and  Johns-Krull C. M., 2011, ApJ, 729, 83
 
 \item[] Yu, C.,  and Lai, D. 2013, MNRAS, 429, 2748
 
\item[]  Yu, C.,  and Li, H. 2009, ApJ, 702, 75

\item[] Zeldovich, Y.B. 1981,  Proc. Roy. Soc. Lond., A, 374, 299

 \item[]  Zhu, Z., Stone, J.M., Rafikov, R.R.,  and Bai, X. 2014, ApJ, 785,
 article id. 122

\end{harvard}

\end{document}